\documentclass{article}
\usepackage{graphicx}

\def\be{\begin{equation}}
\def\ee{\end{equation}}
\def\bea{\begin{eqnarray}}
\def\eea{\end{eqnarray}}

\title{Empirical constraints on vacuum decay in the stringy landscape}
\author{R. Plaga \\ 
Federal Office for Information Security (BSI), D-53175 Bonn, Germany}
\begin{document}
\maketitle
\begin{abstract}
It is generally considered as self evident that the lifetime of our vacuum in the
landscape of string theory cannot be much shorter than the
current age of the universe. Here I show why this lower
limit is invalid. A certain type of ``parallel
universes'' is a necessary consequence of the string-landscape
dynamics and might well allow us to ``survive'' vacuum decay.
As a consequence our stringy vacuum's lifetime is
empirically unconstrained and could be very short.
\\
Based on this counter-intuitive insight
I propose a novel type of laboratory experiment that
searches for an apparent violation of 
the quantum-mechanical Born rule by gravitational effects
on vacuum decay.
If the lifetime of our vacuum should turn out
to be shorter than 6 $\times$ 10$^{-13}$ seconds
such an experiment is sufficiently sensitive 
to determine its value with state-of-the-art equipment.
%with state-of-the-art equipment, otherwise it could 
%set an experimental upper limit.
\end{abstract}

\tableofcontents

\section{Introduction}
\label{intro}
\newtheorem{thm1}{}
\newtheorem{thm2}{}
% stringy landscape
\subsection{The stringy landscape and its decay}
The set of ideas that is usually designated as 
``landscape of string theory''\cite{linde,susskind} (
abbreviated as ``stringy
landscape'' or ``stringscape''),
has been hailed ``as a 
possible radical change in what we accept as a legitimate
foundation for a physical theory''\cite{weinb}.
\\
This synthesis of string and inflationary theory
describes the global universe as
an eternally inflating ``megaverse'' that is continually producing ``pocket
universes''
by tunnelling events between different vacua\cite{freivogel}.
String theory predicts and describes a huge number 
(typically 10$^{\rm hundreds}$) of vacua. 
We inhabit one 
pocket universe (from now on designated as ``our universe'').
Its vacuum (from now on designated as ``our vacuum'')
has a positive cosmological constant, and as a consequence
our universe is entering a quasi de Sitter state presently.
\\
Within string theory
our vacuum (a ``false vacuum'') must eventually decay to some other 
``true-vacuum'' state that has
a smaller energy density. This true vacuum 
features different low-energy
theories and parameters.
Therefore the true vacuum does not support life,
i.e. humans die upon vacuum decay.
Various
authors explored selected, special decay channels and estimated the respective
vacuum-decay rates (e.g. \cite{frey,ceresole,lindetrivedi}). 
Their results form one of the most
direct string-theoretical prediction of an empirically
accessible parameter.
The predicted lifetimes lie typically between the current 
age (13.6 billion years) and the recurrence time (about
10$^{460}$ years) of our universe. It has been argued informally
(and non-conclusively)
that a summation over {\bf all possible} stringscape decay channels possible might
predict vacuum lifetimes that are much
shorter than the current age of the universe\cite{motl_blog}.
%The standard model - that
%must eventually turn out to be a limiting case of string theory in our pocket universe -
%predicts a lifetime of its vacuum far below the current of the universe for the
%current preferred central values of its parameters \cite{pla06}.

\subsection{Aims and structure of this paper}
\label{aims}
My aim is to 
develop concepts to constrain
the rate of vacuum decay in the stringy landscape
empirically and reliably.
Below I assume that the landscape 
exists in the form currently laid out in the literature -
not as an expression of faith but only 
as a purely heuristic basis for further conclusions and test proposals.
My motto is taken from Joseph Polchinski who paraphrased 
Dirac by writing\cite{polchi}:
\\
{\it One should take serious all solutions of 
one's equations\footnote{I would add: this is not a categorical imperative but
a heuristic suggestion!}.}
\\
The current age of our universe
is widely believed to be a conservative lower limit
on the lifetime of our de Sitter vacuum (e.g. \cite{lindetrivedi,denef}).
In section \ref{error} I explain why - contrary to what
common sense might suggest at first sight - this belief is wrong
in the stringy landscape.
%it is more difficult to
%obtain a dependable lower limit on its lifetime than is usually thought.
Sections \ref{gr}
(with appendix \ref{appendix1} that
presents an auxiliary calculation) and \ref{born}, contain the main result of this paper,
a proposal for an experimental procedure to determine either the value
of the our vacuum's decay rate or to place a firm upper limit to it.
Based on the insight that vacuum decay might be extremely
fast, section \ref{limit} estimates the maximal mean lifetime of the
vacuum that could still be determined in a laboratory experiment with
current technology.
%determine a firm lower limit on the
%lifetime of our vacuum or determine the rate of its decay.
%In section \ref{exploit}
%I briefly explore whether vacuum decay can be exploited
%to solve hard computational problems. 
Section \ref{concl} concludes.

\section{
The lifetime of our vacuum 
can be much smaller than the current age of the universe
in the stringy landscape}
\label{error}
%\subsection{Decay of the string-theoretic vacuum}
After reviewing the concept ``parallel universe'' (section \ref{par}) and 
its inevitability in the stringscape (section \ref{inev}), I demonstrate in subsection \ref{surv} why
the (admittedly counter-intuitive) title of this section is true.

\subsection{Overview - What are ``parallel universes''?}
\label{par}
Parallel universes are defined as 
an infinity of distinct universes
that are completely identical to ours 
until a random decision 
makes subsets of them different in the random-decision results.
In a review of this concept Tegmark\cite{tegmark} distinguished
between different types (``levels'') of parallel universes.
``Level I'' denote universes parallel in Minkowski space
whereas e.g. ``level III'' universes coexist in Hilbert space.
The next
subsection \ref{inev} reviews why the existence of
``level I'' parallel universes is inevitable
in the stringy landscape. 
Other types of parallel universes,
like the quantum-mechanical ``many worlds'' (``level III''\cite{tegmark}) are not considered in 
this paper\footnote{Their
existence would strengthen the conclusion of subsection $\ref{surv}$. However, it is not
yet clear whether they are a necessary element of the stringy landscape\cite{suss06,polchi}}.
\\
%It is generally assumed that the lifetime of our 
%vacuum must exceed the current age of the universe(e.g. \cite{lindetrivedi,denef}). 

\subsection{Level I parallel universes (a.k.a. ``parallel Hubble volumes'')
are inevitable in the stringy landscape}
\label{inev}
Coleman \& de Luccia\cite{coleman} showed that pocket universes 
born in a tunnelling event are negatively
curved i.e. they are spatially infinite. Pocket universes 
are formed in this way and are therefore our universe is
spatially infinite in the stringy landscape.
\\
There must be parallel universes in
a spatially infinite universe \cite{brundrit,gar_vil}.
An infinite pocket universe contains
a countably infinite ensemble (
i.e. a set of power $\aleph_0$\cite{cantor}) 
of ``Hubble volumes''\cite{north}. ``Hubble volumes'' 
are defined here as the interior of
light cones that extend back from spacetime points
in Minkowski spacetime at the present age of the universe
to the ``time of recombination'', before which the universe was opaque
to electromagnetic radiation. Because the number of possible histories in a Hubble
volume is finite there must be an infinite number
of Hubble volumes with completely identical histories up to the present, that need
not be identical in the future\cite{gar_vil}. In the following I will
call these: ``parallel Hubble volumes''. 

\subsection{We survive rapid vacuum decay if parallel universes exist}
\label{surv}
Here I demonstrate that the existence of parallel universes
allows us to ``survive'' vacuum decay.
I recently argued\cite{pla06} that 
the Standard-Model
vacuum could have a much shorter lifetime than 
our (pocket) universe's age, 
if parallel universes 
exist\footnote{I generalized the ``quantum suicide'' concept\cite{squires},
abstracting the ``suicide'' and extending it to all types of parallel universes.
The fact that the existence of parallel universes allow us to survive vacuum decay
was first pointed out in a science-fiction short story\cite{gribbin}.}. 
Here I generalize this argument and apply it to the stringy
landscape.
\\ 
It is well known that vacuum tunnelling events are exponentially
unlikely to take place ``all over the universe''\cite{linde}. Rather they are
typically seeded in ``critical bubbles'' which finite size depends 
on details of the transition\cite{coleman,linde}. This bubble then expands with a velocity
that I will assume to be very close\footnote{To a fractional precision of
at least 10$^{-20}$. This condition is not necessarily 
but possibly fulfilled in a vacuum transition (it is e.g. for the decay of
the standard model (SM) vacuum)\cite{pla06}.} or equal to c in this paper.
\\
Let ``q'' be the probability that there is no critical bubble in 
a given Hubble volume, i.e. vacuum decay did not 
``took place'', yet. For vacuum lifetimes that
are much shorter than the present age of the universe, q can be extremely
small, but it always remains finite.
The number of ``parallel universes'' since the big bang is reduced by vacuum decay
from a number of $\aleph_0$ to $\aleph_0$ $\times$ q Hubble volumes. This ``reduction process''
cannot be perceived by human observers, because, under
the assumptions stated above, relativistic causality prevents any 
indication of an impending decay to reach her on a time scale longer
than the minimum one necessary for becoming conscious 
of the impending doom\cite{pla06}.
\\
A basic assumption of the stringy landscape is that the perceived properties our 
Hubble volume do
not have to be likely. Anthropic 
reasoning asserts that experience only requires that Hubble volumes with
such properties exist somewhere in our pocket universe.
In the same sense the stability of our vacuum does not need to be likely.
It is enough if un-decayed parallel Hubble volumes
exist somewhere in our
universe.
In other words:
From the fact that we perceive our vacuum 
to be stable we can only infer that stable parallel Hubble volumes exist somewhere in 
our pocket universe.
\\
The relation
\begin{equation}
\aleph_0 = \aleph_0 q
\end{equation}
is true for all finite q\cite{cantor}.
Vacuum decay, no matter how fast, can therefore not
reduce the number of $\aleph_0$ parallel Hubble volumes.
Because $\aleph_0$ stable Hubble volumes do exist 
in our pocket universe even in the presence of
arbitrarily fast vacuum decay,
one cannot derive any lower limit on the lifetime of our stringy de Sitter vacuum
from the mere fact ``that we still exist''. In reality vacuum decay could be
extremely fast.

\section{How to experimentally determine or constrain the lifetime
of our vacuum} 
\label{howto}
%skimmer help
Subsections (\ref{gr}, \ref{born}) propose that the decay probability 
of a quantum mechanical system can be made to 
depend on the decay rate of the vacuum in a
suitably designed laboratory experiment.
Subsection \ref{limit} asserts that such an
experiment might be sensitive enough to determine
a sufficiently large vacuum-decay rate with 
state-of-the-art methods.
%the experimentally accessible effective probability
%``p$_e$'', that depends on the
%lifetime of our vacuum.
\subsection{How to slow down vacuum decay in the laboratory}
\label{gr}
Here I calculate how many vacuum-decay events ``$\Delta$n'' less occur at a given
point in space within 
a time period $\Delta$t as the consequence of ``blowing up'' a massive sphere mechanically.
\begin{figure}
%\rule{5cm}{0.2mm}\hfill\rule{5cm}{0.2mm}
%\vskip 2.5cm
%\rule{5cm}{0.2mm}\hfill\rule{5cm}{0.2mm}
%\epsfig{figure=figu2.eps,height=3in}
\includegraphics[angle=0,totalheight=4.5in]{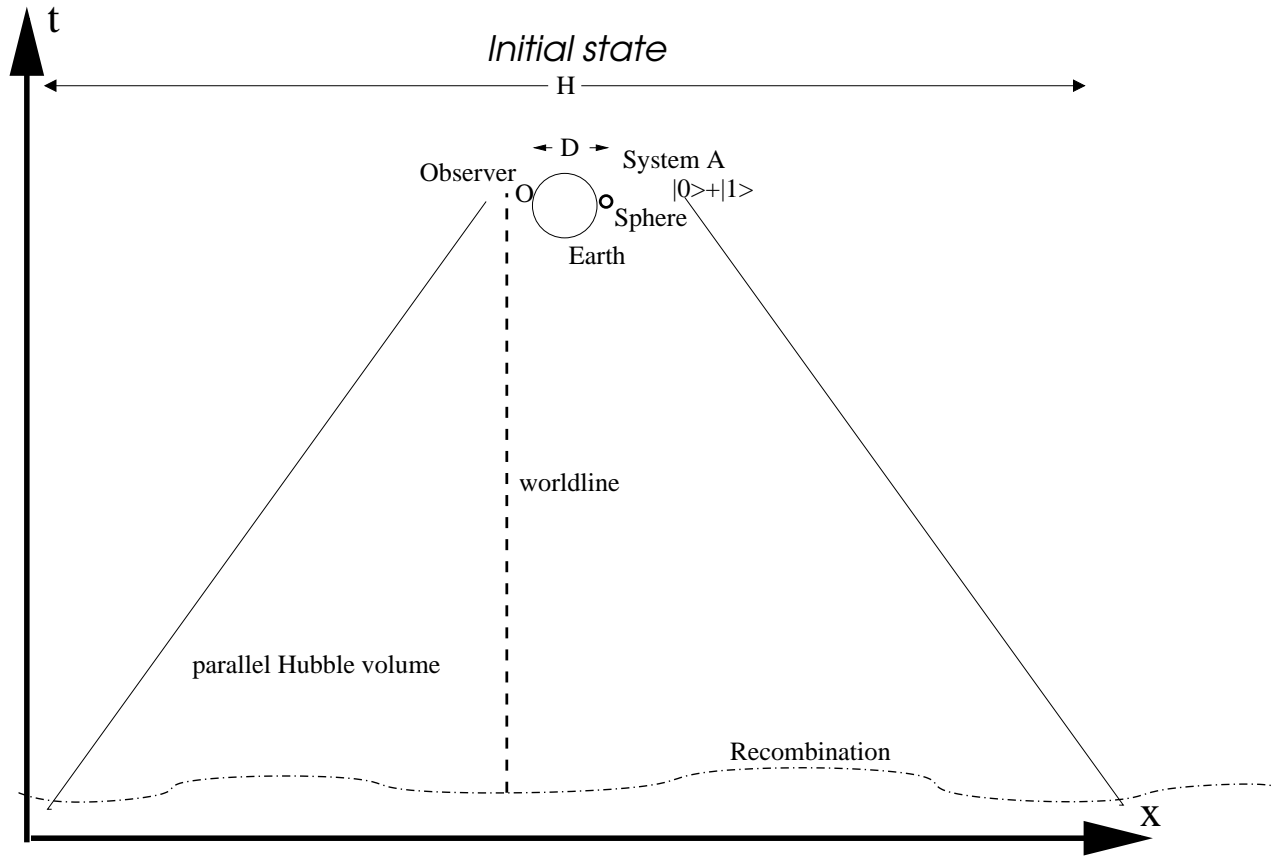}
% pdf version {stringy_pic14_initial2.pdf}
%\epsfbox
\caption{
Illustration of the proposed setup, described in section \ref{born}, in the
initial state, before the measurement is performed (assuming that vacuum decay
has not taken place).
The space-time diagram  depicts 
part of our pocket universe.
The full cone-shaped lines symbolize - not to scale - 
the spatial confines of
a parallel Hubble volume that
extends back from the location of the experiment.
Today one can look back to a distance 
of H $\approx$ 10$^{10}$ light
years. The lower dot-dashed line delineates the time of recombination
(380000 years after the big bang), the universe is transparent to light
only above the line.
The small circle with a diameter of D $\approx$ 10$^4$ km stands for 
the earth today (13.6 billion years after
the big bang). The mechanical sphere and the measured system A
are on one side of the earth, 
and the observer's worldline - marked by the thick dashed line - on
the other. The quantum state of A is indicated here and
in fig.\ref{explot}.
\label{exploti}}
\end{figure}
\begin{figure}
%\rule{5cm}{0.2mm}\hfill\rule{5cm}{0.2mm}
%\vskip 2.5cm
%\rule{5cm}{0.2mm}\hfill\rule{5cm}{0.2mm}
%\epsfig{figure=figu2.eps,height=3in}
\includegraphics[angle=0,totalheight=4.5in]{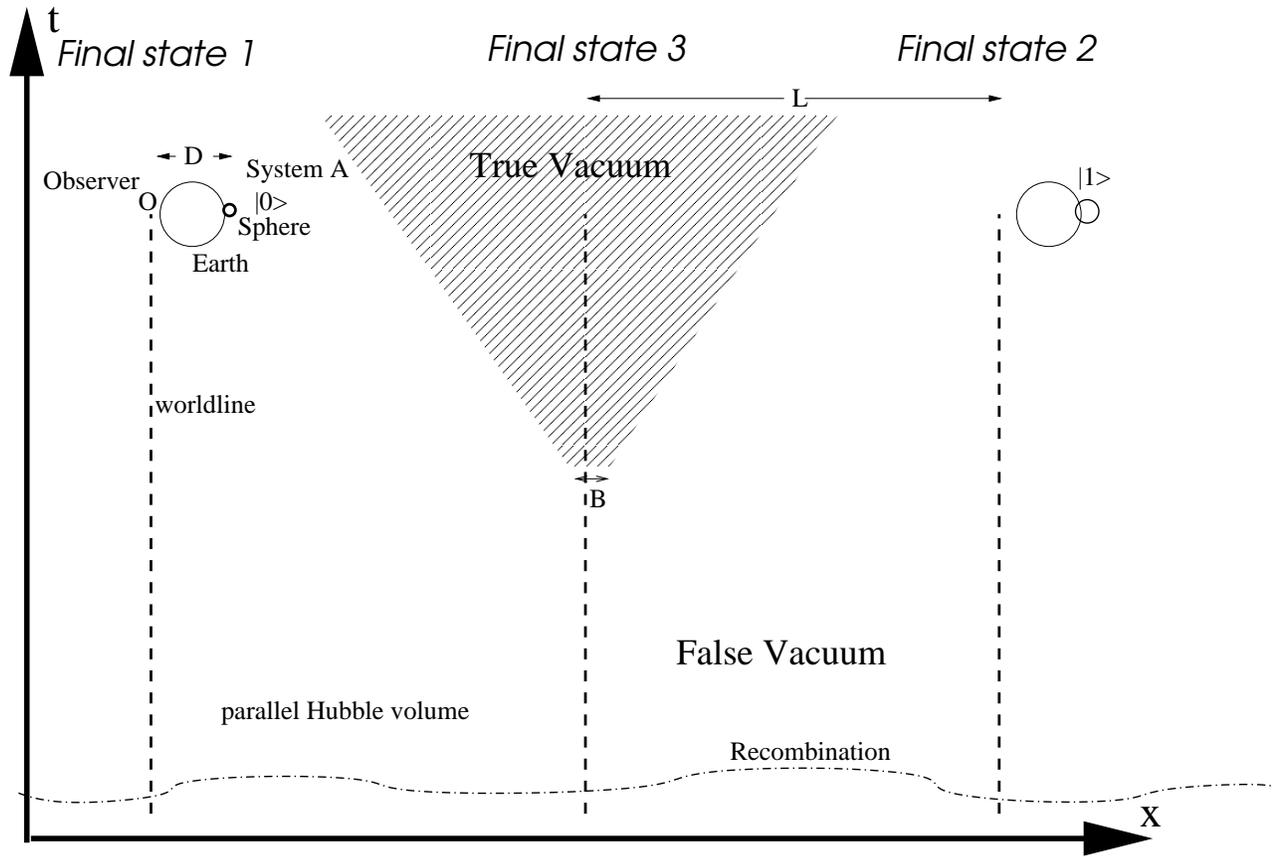}
%{stringy_pic14_final2.pdf}
%\epsfbox
\caption{
Illustration of the experiment
in its three possible final states after a just
performed experiment (that is described in subsection \ref{exact}).
The distance between the location
of three final states in parallel Hubble volumes has
been estimated to L $\approx$ 10$^{{10}^{29}}$ m\cite{tegmark}.
The sphere is inflated in final state 2
upon the measurement result ``1'' of system A.
For final state 3 the part of spacetime
in the true-vacuum state is hatched. The vacuum transition began in a bubble of
diameter ``B''. The value of B depends on the precise nature of the transition. 
E.g. for a SM transition 
B $\approx$ 10$^{-33}$ m.
\label{explot}}
\end{figure}
The expected number of vacuum-decay events ``$n$'' 
that affect\footnote{``Affect'' means that the
transition front of a vacuum decay event eventually reaches P.} 
a given point P in space within the time
period ``$\Delta$t''
is given as:
\begin{equation}
n = \rho \times (4/3 \pi c^3 \times (\Delta t)^4 - \Delta V_4).
\label{n_dec}
\end{equation}
Here $\rho$
is the density of ``critical vacuum-decay bubbles'' (see section
\ref{surv}) in 4-space.
% and t$_{\rm vac}$ is the mean lifetime of the vacuum.
$\Delta$V$_4$ is a small general-relativistic correction 
that depends on the distribution of masses in space (see below).
The mean lifetime of our vacuum ``t$_{\rm vac}$'' is defined as
\begin{equation}
t_{\rm vac} = \left( 1 \over {\rho c^3} \right)^{1\over 4}
\label{tvac}
\end{equation}
\\
Let us evaluate $\Delta$ V$_4$ 
for a sphere centred at a point P and
filled with a perfect fluid of constant density
with an initial radius r$_{f1}$ and empty space
beyond it. It can then be shown within general relativity (GR)\cite{gr} (see appendix \ref{appendix1}) that:
\begin{equation}
\Delta V_4(r_{f1}) = - {2 \over 5} \pi r_{S} \Delta t  r_{f1}^2.
\label{f1}
\end{equation}
Here r$_S$ = 2 G m/c$^2$ is the Schwarzschild radius of the sphere, m its
mass and G Newton's constant of gravitation.
If the sphere is ``mechanically blown up'' (i.e. its radius is
increased, while holding its mass constant thereby decreasing the density
by the cube of the radius-increase factor) 
to a larger radius r$_{f2}$ one obtains:
\begin{equation}
\Delta V_4(r_{f2})  = - {2 \over 5} \pi r_{S} \Delta t  r_{f2}^2.
\label{f2}
\end{equation}
The ``blow up`'' thus ``destroys'' a volume of 4-space\footnote{
This result is intuitive:
Near a massive sphere
space is dilated in the radial direction and time is compressed.
Outside the sphere the two effects cancel, so
the 4-volume is exactly the same as in Minkowski space.
Within a filled perfect-fluid sphere time is compressed to a 
greater degree so that the spacetime volume within
the sphere becomes smaller upon inflation.
The ``destruction'' of spacetime is mainly due to general-relativistic
time dilation. There are fewer vacuum-decay events upon
inflation because ``time ticks
slower within a massive sphere''.}:
\begin{equation}
\Delta \Delta V_4  = {2 \over 5} \pi r_{S} \Delta t  (r_{f2}^2-r_{f1}^2).
\label{deldel}
\end{equation}
The blow up of the sphere 
therefore decreases the total number of vacuum decays during a time $\Delta$ t by:
\begin{equation}
\Delta n  = \rho \times \Delta \Delta V_4
\label{deln}
\end{equation}
decays. Vacuum decay can be slowed down by changing the distribution
of mass in space.

\subsection{
Proposal for an experiment that probes vacuum stability via apparent violations of the Born rule}
\label{born}
This section proposes an experimental setup to 
determine the effect of a reduced number of vacuum-decay
events.
A schematic picture of the initial and final state
of this experiment is provided in figs. \ref{exploti} and \ref{explot}, respectively.
\subsubsection{Detailed description of the experiment}
\label{exact}
At a time ``t$_m$'' (the origin t=0 is set to the recombination time) 
the state of a quantum-mechanical system A is measured
(see fig.\ref{explot} for the initial state). 
With a certain probability ``1-p'', that
is given be the Born rule, the outcome be "0", with the probability p
it be ``1''. If, and only if, the outcome is ``1'', a massive sphere that
is independent of A but located at the same position in space, is
mechanically and
automatically ``blown up'' (defined in the previous section
\ref{gr}). The blow up is to be concluded within a time period $\Delta$t$_{bu}$.
An inertial observer ``O'' is located at a large enough distance d from the sphere, 
so that d/c $\gg$ $\Delta$t$_{bu}$.
%E.g. if observer O and system A are at the opposite sides of the earth, the
%blow up must proceed on a time scale smaller than 0.03 seconds.
\\
The earliest time that O can measure the outcome of the experiment
is a time t$_o$ $\approx$ t$_m$ + d/c. To avoid (bound to be controversial) questions
concerning the nature of quantum-mechanical measurements, I 
assume that O's measurement takes place instantaneously at time
t$_o$. This is the most conservative possible option, because - as we
will see below\footnote{The value of $\Delta$t in subsection \ref{limit} would be longer
under different assumptions,and this would decrease $t_{\rm vac}^{\rm crit}$.} - it is under
this assumption that the proposed experiment
leads to the least restrictive upper limits on the vacuum-decay rate.  
\subsubsection{Determination of the effective probability p$_e$}
\label{pe}
My aim is to calculate the ``{\bf e}ffective'' probability ``p$_e$''
(rather than p) with which observer O 
finds the outcome ``1'' due to the 
effect of vacuum decay. The basic idea is that there are 
``more'' parallel universes with
outcome ``1'' because their vacuum-decay proceeds slower according to 
the arguments in section
\ref{gr}. Intuitively this should mean that the p$_e$ 
increases with the vacuum-decay rate. But we run into two
problems:
\\
Firstly, according to the argument
of subsection \ref{surv}, once a quantum mechanical measurement is performed,
the vacuum decay rate can have no more effect on its outcome probability. 
We addressed this 
problem by delaying the earliest time at which
the system can be measured by $\Delta$t $\approx$ d/c 
in the experimental design.
At least during this time period
the outcome-dependent vacuum-decay rates can ``take effect''.
\\
Secondly there is an infinite number of parallel Hubble volumes.
Therefore the ratio of outcomes ``0'' and ``1'' is 
a quotient of infinite numbers and therefore ill defined.
A gauge-invariant regulator for the countably infinite number of 
Hubble volumes is needed. This problem is closely related to the
problem of counting pocket universes with different vacuum types
in the megaverse\cite{vile06}. 
A method proposed by Easther et al.\cite{easther}
to attack this problem can be readily adapted to the present case.
Easther et al. consider a collection
of a finite number N of initial points. In our case these
points are chosen at the time of recombination in N different
``parallel Hubble volumes'' in such a way that their
world lines pass up to the location of observer O at time t$_o$.
\\
These world lines can end in three final states (see fig. \ref{explot}):
\\
1.  $|$our vacuum, outcome 0$>$
\\
2.  $|$our vacuum, outcome 1, sphere blown up$>$
\\
3. $|$true vacuum, no observer$>$, in which our vacuum has
decayed, and no human life is possible\footnote{In this case the
worldline passes up to the position where the observer O would have
been had the vacuum not decayed.}. 
\\
Easther et al.\cite{easther} propose to estimate p$_e$
as the ratio of worldlines that end in state 2 to the
total number of world lines that support ``observers''.
They further argue
anthropically (and in full accordance with the
argument of our section \ref{surv}) that
final state 3, in which no human observers exist,
must be, quote, ``dropped from consideration''.
\\
The number of world lines ending in state 1 is given as:
\begin{equation}
N_1= N \times P_{d1} \times (1-p)
\label{n1}
\end{equation} 
Here P$_{d1}$ is the probability for no vacuum decay on the world
line within the time period t$_0$ $\approx$ 13.6 billion years since the big bang:
\begin{equation}
P_{d1} = e^{- {\rm n_0}} 
\label{d1}
\end{equation} 
Here n$_0$ is the expected value of vacuum decay events n (eq.\ref{n_dec})
for $\Delta$t=t$_0$.
The number of world lines ending in state 2 is given as:
\begin{equation}
N_2= N \times P_{d2} \times p 
\label{n2}
\end{equation} 
The probability for vacuum decay is smaller for this case
due to the smaller amount of 4-space from
the ``blow up'' of the sphere ($\Delta$n is given in eq.(\ref{deln})).
\begin{equation}
P_{d2} = P_{d1} \times e^{\Delta n}
\label{d2}
\end{equation} 
Finally 
\begin{equation}
N_3 = N - N_1 - N_2 
\end{equation}
The probability p$_e$ to find result ``1'' in a decaying vacuum
can then be derived from eqs.(\ref{d1},\ref{d2})  as 
(analogous to eq. (4.2) in Easther  et al.\cite{easther})
\begin{equation}
p_e= {N_2 \over {N_1 + N_2}} = {{p  e^{\Delta n}} \over {(1-p)+p e^{\Delta n}}}
\label{pfin}
\end{equation}
and the effective probability for result ``0'' is
1 -- p$_e$. For a stable vacuum $\Delta$n = 0, and one
gets the usual result p$_e$ = p. If $\Delta$n is finite,
p$_e$ is larger than p.
In this case the lifetime of the
vacuum t$_{\rm vac}$ can be inferred from eqs.(\ref{pfin},\ref{tvac},\ref{deln}).
The implied violation of the Born rule is only apparent because
the increased value of p$_e$ is a subjective
consequence of the increased probability of survival to vacuum decay
when result ``1'' was obtained.

\subsection{Sensitivity of the experiment - achievable limit on the vacuum-decay lifetime}
\label{limit}
I roughly estimate the sensitivity of a reasonably sized experiment as described in 
the previous subsection \ref{born}.
\\
Upon result ``1'' a sphere with a mass of 1000 kg be blown up
from a radius of 0.5 m to 10 m within $\Delta$t$_{bu}$ = 5 msecs. This could be approximated 
e.g. by a chemical explosion of 1 kt of TNT - within air\footnote{The 
velocity of TNT debris is estimated $\approx$ 2 km/sec until it is slowed
down to much smaller velocities at $\approx$ 10 m due to the swept-up air mass.}.
Observer O is located at a distance d=10000 km (corresponding to
two distant sites on earth) from the sphere so that d/c = 30 msec and vacuum
can ``take effect'' (see second paragraph of subsection \ref{pe}) for
$\Delta$t=25 msec.
Inserting these values into eq.(\ref{deldel})
one finds via eqs.(\ref{tvac},\ref{deln}) that
$\Delta$n becomes $>$ 1 for a ``critical'' mean 
lifetime of our vacuum t$_{\rm vac}^{\rm crit}$ 
of:
\begin{equation}
t_{\rm vac}^{\rm crit} < 6 \times 10^{-13} {\rm seconds}.
\label{crit}
\end{equation}
Should such an experiment find that p$_e$ (eq.(\ref{pfin})) is equal to p
within errors the lifetime of the vacuum must be  $>$ 6 $\times$ 10$^{-13}$ seconds.
Shorter lifetimes increase p$_e$. 
%Its experimentally determined value can be used
%to determine the mean lifetime of the vacuum t$_{\rm vac}$ by
%reverting the derivation of eq.(\ref{crit}). 
I stress again the {\bf counter-intuitive fact} that such short lifetime are {\bf not} empirically
ruled by the fact that we are ``still alive'' (see section \ref{surv}).
\\
While 10$^{-13}$ sec is about 30 orders of magnitude smaller
than the current age of the universe, it is still about 31 orders of
magnitude larger then the Planck time, so there is a lot of ``room
at the bottom'' for possibly ``detectable'' lifetimes of our vacuum.

\section{Conclusion and outlook}
\label{concl}
Because the landscape firmly predicts an infinity of ``parallel
Hubble volumes'', some ``copies'' of our world will always
survive, even if the probability to survive
vacuum decay in any individual world
is arbitrarily small. Experience only tells us that at least
one surviving copy exists, and cannot be the basis
for an empirical limit on the mean lifetime of the vacuum.
{\bf Therefore it is definitely incorrect to consider the current age of the universe
as an upper limit to the lifetime of our vacuum within the 
stringy landscape.}
\\
I proposed a novel type of laboratory experiment to set a
less restrictive but more accurate upper limit to
the lifetime of our vacuum.
If the mean lifetime of the vacuum is shorter than about
0.6 picoseconds (and such a short lifetime is {\bf not} empirically
ruled out) vacuum-decay could be detected with a technologically
feasible experiment.
\\
%The proposed experiment for obtaining empirical evidence about
%the decay lifetime will probably 
%yield a upper limit on t$_{\rm vac}^{\rm crit}$ . 
The proposed experiment is unlikely to be the ``last word'' on this
problem. It is not fully understood how
to calculate probabilities or how to use
transfinite numbers in the landscape, yet.
But one conclusion seems fairly definite: 
the lifetime of our vacuum - a quantity calculable  
with string theory (even if a calculation
of the total decay rate is still outstanding) -
can be subjected to a direct laboratory test.
\\
A detection of vacuum decay would lend support to crucial concepts of 
the stringy landscape like
%I do not mean the one of fast vacuum decay (a phenomenon
%that, after all, can even occur in the Standard Model) but  
the existence of parallel universes and
the correctness of anthropic reasoning.
It seems of course more likely that the experiment proposed in section \ref{born}
will merely yield an upper limit to the decay rate of our stringy vacuum. But
even this will be a genuine experimental
constraint on a parameter of the stringy landscape.
\\
Polchinski's motto, quoted in section \ref{aims} urges us to take 
the 10$^{\rm hundreds}$ vacuum solutions of string theory serious. This paper
argues to do the same with the $\aleph_0$ Hubble volumes in our pocket universe.

%\section*{Acknowledgements}
%I sincerely thank C.L.S.Carozze, R.A.Faolutte and E.R.Elinina
%for penetrating criticism of previous versions of this
%manuscript.
%Scott Aaronson, Kari Enqvist and Gino Isidori patiently answered my
%questions concerning computational complexity and vacuum instability.
%Erich Joos, Dennis K\"ugler and Alvaro de Rujula constructively 
%criticised earlier versions of this manuscript.
%I thank them all.It is based on the manipulation 

%\section*{References}

\section{Appendix - The 4-volume of perfect-fluid sphere}

\label{appendix1}
The aim of this appendix is to calculate the 4-volume of a perfect-fluid
sphere in general relativity to derive eq.(\ref{f1}).
Consider a incompressible-fluid sphere.
Both in the interior and to the exterior of such a sphere
only the diagonal components of the metric tensor 
are non zero. In the interior the metric components are\cite{schwarzschild}
(I use a spacelike ($-$,+,+,+) metric):
\begin{eqnarray}
g_{tt} & = & {{{1}\over{2}} \left({3 \sqrt{1-r_S/r_f}-\sqrt{1-{{r_S r^2}\over{r_f^3}}}}\right)^2} 
\nonumber
\\
g_{rr} & = & -{\left(1-{r_S r^2}\over{r_f^3}\right)}^{-1} 
\nonumber
\\
g_{\theta\theta} & = & -r^2 
\nonumber
\\
g_{\phi\phi} & = & -(r sin({\theta}))^2
\label{int_metric}
\end{eqnarray}
r,$\theta$,$\phi$ are the usual polar coordinates.
r$_f$ is the radius of the sphere,
m is the total mass of the
sphere, r$_S$ is the Schwarzschild radius (2 G m)/c$^2$ and G is the constant of
gravitation.
The space-time volume V$_4$ of this 3-sphere extended in time from t$_1$ to t$_2$
is the desired result:
\begin{eqnarray}
V_4 = \int^{t_2}_{t_1} \int^{r_f}_0 \int^{-\pi}_{\pi} \int^{2\pi}_0 
\sqrt{-det(g)} \ {dt} \ {dr} \ {d\theta} \ {d\phi} = ({4 \over 3} \pi r_f^3 -
\nonumber
\\
{2 \over 5} \pi r_f^2 r_S) \Delta t
\label{vol}
\end{eqnarray}
Here 
$\Delta$t=t$_2$-t$_1$.
Because r$_S$/r$_f$ $\ll$ 1 in all laboratory situations,
terms of order 2 and higher in this quotient were neglected for the
final expression.
At the exterior to the sphere the diagonal components 
of the metric components  
are\cite{schwarzschild}:
\begin{eqnarray}
g_{tt} & = & (1-r_S/r_f) 
\nonumber
\\
g_{rr} & = & {(1-r_S/r_f)}^{-1} 
\nonumber
\\
g_{\theta\theta} & = & -r^2 
\nonumber
\\
g_{\phi\phi} & = & -(r sin({\theta}))^2
\label{ext_metric}
\end{eqnarray}
Because here
g$_{rr}$=1/g$_{tt}$ , det($-$g) (and therefore
the space-time volume exterior to the
sphere) is identical to the one in flat space time.

\begin{thebibliography}{99}
\bibitem{linde}
A. Linde, Particle physics and inflationary cosmology, Harwood, Chur, 1990.
\bibitem{susskind}
L. Susskind, {\it The anthropic landscape of
string theory}, hep-th/0302219v1.
\bibitem{weinb} S. Weinberg, {\it Living in the Multiverse}, hep-th/0511037v1.
\bibitem{freivogel}
B. Freivogel, M. Kleban, M. Rodriguez Martinez, L. Susskind, 
{\it Observational Consequences of a Landscape}, hep-th/0505232v2.
\bibitem{frey}
A. R. Frey, M. Lippert, B. Williams, {\it The fall of stringy de Sitter},
Phys.Rev. D68 (2003) 046008.
\bibitem{ceresole}
A. Ceresole, G. Dall' Agata, A. Giryavets, R. Kallosh, A. Linde,
{\it Domain walls, near-BPS bubbles and probabilities in the
landscape} hep-th/0605266v2.
\bibitem{lindetrivedi}
S. Kachru, R. Kallosh, A.Linde, S.P. Trivedi, {\it de Sitter vacua in string theory}
Phys.Rev. D68 (2003) 046005.
\bibitem{motl_blog}
L. Motl, {\it Landscape decay channels}, motls.blogspot.com/2005/11/landscape-decay-channels.html,
2005.
\bibitem{polchi} J. Polchinski, {\it The Cosmological Constant And the String
Landscape}, hep-th/0603249v2.
%\bibitem{starobinsky99} A.A. Starobinsky, {\it Future and origin of our universe: modern view}, 
%Grav. Cosmol. 6 (2000) 157.
\bibitem{denef}
F. Denef, M.R. Douglas, {\it Computational complexity of the landscape I},
hep-th/0602072v2.
\bibitem{tegmark}
M. Tegmark, {\it Parallel universes}, In: {\it Science
and the Ultimate Reality: From Quantum to Cosmos}, J.D.Barrow, P.C.W. Davies, 
C.L. Harper eds., Cambridge University Press, Cambridge, 2003;
astro-ph/0302131v1.
\bibitem{suss06} L.Susskind, priv. comm. (2006).
\bibitem{coleman} 
S.R. Coleman, F. DeLuccia {\it Gravitational Effects On and Of Vacuum Decay}
Phys.Rev.D 21 (1980) 3305.
%\bibitem{goheer} N. Goheer, M. Kleban, L. Susskind, {\it The Trouble with de Sitter Space}, JHEP 7 (2003) 056.
\bibitem{brundrit} Ellis G.F.R., Brundrit G.B., 1979, {\it Life in the Infinite Universe},
Q.Jl.R.astr.Soc. {\bf 20},37-41.
\bibitem{gar_vil}
Garriga J., Vilenkin A., 2001, {\it Many worlds in one},
Phys.Rev. {\bf D64},043511; gr-qc/0102010.
\bibitem{cantor}
G. Cantor, 1895, {\it Beitr\"age zur Begr\"undung der transfiniten Mengenlehre}, Math.Ann. 46,481
\bibitem{north}
J.D. North, The measure of the universe, Dover, New York, 1999; Chapt. 17.
\bibitem{pla06}
R. Plaga, {\it New physics beyond the standard model of
particle physics and parallel universes}, Phys.Lett. B634 (2006) 116.
\bibitem{squires} E. Squires, The Mystery of the Quantum World, Institute
of Physics Publishing, Bristol, 1986.
\bibitem{gribbin} J. Gribbin, {\it The doomsday device}, Analog 105 (1985) 120.
\bibitem{gr} 
Einstein A., 1956, {\it Grundz\"uge der Relativit\"atstheorie}, (Vieweg 
\& Sohn, Braunschweig).
\bibitem{vile06}
A. Vilenkin, {\it Probabilities in the landscape}, hep-th/0602264v2.
\bibitem{easther}
R. Easther, E.A. Lim, M.R. Martin, {\it Counting pockets with
World Lines in Eternal Inflation}, astro-ph/0511233v3.
%\bibitem{moravec}
%Moravec H., 1988, Mind children. The future of Robot and
%Human Intelligence, Harvard University Press, Cambridge; Appendix 3. 
%\bibitem{aes} csrc.nist.gov/publications/fips/fips197/fips-197.pdf 
\bibitem{schwarzschild} Schwarzschild K., 1916, {\it \"Uber das Gravitationsfeld einer Kugel aus
inkompressibler Fl\"ussigkeit nach der Einsteinschen Theorie}, Sitzungber. Preu\ss.
Akad. Wiss. Berlin, Kl. Math.-Phys. Tech. {\bf 18}, 424-434.
\end{thebibliography}
\end{document}